\documentclass[twocolumn]{aastex62}
\usepackage{natbib,amsmath}
\usepackage{xspace}
\usepackage{hyperref} \usepackage{float} \usepackage{subfigure}

\newcommand\ourstar{\mbox{TOI 2134}\xspace}
\newcommand\planet{\mbox{TOI 2134b}\xspace}
\providecommand{\e}[1]{\ensuremath{\times 10^{#1}}}

\graphicspath{{./}{figures/}}

\begin{document}

\title{Outflowing helium from a mature mini-Neptune}

\correspondingauthor{Michael Zhang}
\email{mzzhang2014@gmail.com}

\author[0000-0002-0659-1783]{Michael Zhang}
\altaffiliation{51 Pegasi b Fellow}
\affil{Department of Astronomy \& Astrophysics \\
University of Chicago\\
Chicago, IL 60637}

\author[0000-0002-8958-0683]{Fei Dai}
\altaffiliation{NASA Sagan Fellow}
\affiliation{Division of Geological and Planetary Sciences,
1200 E California Blvd, Pasadena, CA, 91125, USA}
\affiliation{Department of Astronomy, California Institute of Technology, Pasadena, CA 91125, USA}

\author[0000-0003-4733-6532]{Jacob L.\ Bean}
\affil{Department of Astronomy \& Astrophysics \\
University of Chicago\\
Chicago, IL 60637}

\author[0000-0002-5375-4725]{Heather A. Knutson}
\affiliation{Division of Geological and Planetary Sciences, California Institute of Technology}

\author[0000-0002-0594-7805]{Federica Rescigno}
\affiliation{Department of Physics and Astronomy, University of Exeter, Stocker Rd, Exeter, EX4 4QL, UK}

\begin{abstract}
We announce the detection of escaping helium from \planet, a mini-Neptune a few Gyr old.  The average in-transit absorption spectrum shows a peak of $0.37 \pm 0.05$\% and an equivalent width of $W_{\rm avg}=3.3 \pm 0.3$ m\AA{}.  Among all planets with helium detections, \planet is the only mature mini-Neptune, has the smallest helium signal, and experiences the lowest XUV flux.  Putting \planet in the context of all other helium detections, we report the detection of a strong (p=3.0\e{-5}) and theoretically expected correlation between $F_{\rm XUV}/\rho_{\rm XUV}$ (proportional to the energy-limited mass loss rate) and $R_* W_{\rm avg}$ (roughly proportional to the observationally inferred mass loss rate).  Here, $W_{\rm avg}$ is the equivalent width of the helium absorption and $\rho_{\rm XUV}$ is the density of the planet within the XUV photosphere, but the correlation is similarly strong if we use the optical photosphere.  \planet anchors the relation, having the lowest value on both axes.  We encourage further observations to fill in missing regions of this parameter space and improve estimates of $F_{\rm XUV}$.
\end{abstract}


\section{Introduction} 
\label{sec:intro}
Atmospheric escape fundamentally shapes the properties of exoplanets.  It likely carves the radius gap that separates the small, dense super-Earths from the larger and puffier mini-Neptunes \citep{fulton_2017,fulton_2018}, either through photoevaporation by stellar XUV \citep{lopez_2013, owen_2017, mills_2017}, or by core-powered mass loss powered by the planet's own cooling luminosity \citep{ginzburg_2018,gupta_2019}.  Among terrestrial planets, atmospheric escape has momentous implications for habitability: planets that have lost their atmospheres are unlikely to have liquid water on their surfaces.

The first escaping atmosphere was detected in Ly$\alpha$ absorption 20 years ago \citep{vidal-madjar_2003}, but only a handful of other Ly$\alpha$ detections have followed.  The year 2018 saw the first successful use of an alternate mass loss probe: the 1083 nm transition between the metastable triplet ground state and a triplet excited state \citep{spake_2022}.  Only a few helium atoms per million are in the triplet ground state in the best of circumstances, and not all stellar types are equally effective at population this state \citep{oklopcic_2019}.  Nevertheless, the accessibility of the line from the ground, the copious stellar photons at this wavelength, and the lack of interstellar extinction more than make up for these downsides, and more than a dozen outflows have been definitively detected in this line \citep{dos_santos_2022}.

Recently, we detected the first helium outflow from a young mini-Neptune \citep{zhang_2022c}, followed by detections from three other young mini-Neptunes \citep{zhang_2023}.  The widths of the helium absorption signals suggest a photoevaporative outflow while disfavoring the core-powered mass loss scenario, and the equivalent widths imply mass loss rates sufficient to strip a substantial fraction of the atmosphere on Gyr timescales.  These observations are important for testing mass loss models, which suffer from large theoretical uncertainties.  For example, mini-Neptunes may have very high metallicity atmospheres \citep{kempton_2023}, which have lower hydrogen/helium abundance, slower outflows (because of the higher mean molecular weight), and higher temperatures, suppressing the outflow \citep{zhang_2022c}.  Planetary magnetic fields can affect mass loss in complex ways (e.g. \citealt{schreyer_2023, ramstad_2021}). Even with strong mass loss, it is possible for the atmosphere to be replenished by outgassing, for example from hydrogen and water dissolved in magma \citep{chachan_2018,kite_2020}.   The large uncertainties in all of these processes make atmospheric escape difficult to model.  It is therefore important to catch mini-Neptunes of different ages and irradiation levels in the process of losing their envelopes, in order to have observational data to nail down theoretical models.

\begin{table}[ht]
    \caption{Summary of system properties from \cite{rescigno_2023}}
    \centering
    \setlength{\tabcolsep}{6pt}
    \begin{tabular}{C C}
        \hline
        \text{Property} & \text{Value}\\
	\hline 
        R_* (R_\Sun) & 0.709 \pm 0.017 \\
        M_* (M_\Sun) & 0.744 \pm 0.027 \\
        T_{\rm eff} (K) & 4580 \pm 54 \\
        log(g) & 4.8 \pm 0.3 \\
        \text{[Fe/H]} & 0.12 \pm 0.02 \\
        P(d) & 9.2292004 \pm 6.3\e{-6} \\
        R_p/R_* & 0.03475 \pm 0.00034 \\
        R_p (R_\Earth) & 2.69 \pm 0.16 \\
        a/R_* & 23.66 \pm 0.52 \\
        a (\text{AU}) & 0.078 \pm 0.0009 \\
        b & 0.20 \pm 0.12\\
        e & 0.06_{-0.04}^{+0.04}\\
        T_{eq} & 666 \pm 8 \\
        M_p (M_\Earth) & 9.13_{-0.76}^{+0.78} \\
        D (pc) & 22.655 \pm 0.007 \\
        \hline
    \end{tabular}
    \label{table:system_properties}
\end{table}

In this paper, we present the first detection of escaping helium from \planet, a warm mini-Neptune orbiting a nearby (23 pc) X-ray-quiet K dwarf \citep{rescigno_2023}.  Although we initially targeted it as part of our program to observe young mini-Neptunes \citep{zhang_2023}, additional data have shown that it is a mature planet ($\sim$2 Gyr).  For convenience, Table \ref{table:system_properties} presents relevant stellar and planetary properties.  We describe the observations and reduction in Section \ref{sec:observations} and the helium outflow's properties in Section \ref{sec:results}, before comparing \planet to other helium detections in Section \ref{sec:discussion} and concluding in Section \ref{sec:conclusion}.

\section{Observations and Data Reduction}
\label{sec:observations}
Over the past two years, we have been carrying out a survey of escaping helium from young ($\lesssim$1 Gyr) mini-Neptunes orbiting nearby K dwarfs.  The survey, described in \cite{zhang_2023}, uses Keck's high-resolution NIRSPEC spectrograph to detect 1083 nm helium absorption, the TESS light curve to measure the star's rotation period, and XMM-Newton data to measure the star's X-ray spectrum.  The survey has detected helium from all four of its first four targets.  \ourstar was the fifth target to be observed as part of this survey.

We observed \planet with Keck/NIRSPEC from 2022-06-18 09:44 UTC to 14:35 UTC, consisting of 1.3 h of pre-ingress baseline, the 3.0 h transit, and 0.5 h of post-egress baseline.  As usual, we used the 12 x 0.432$\arcsec$ slit, giving us a spectral resolution of 32,000.  Also as usual, we took 60 s exposures in an ABBA nod pattern.  \ourstar is brighter than the four targets in \cite{zhang_2023}, giving us a typical SNR of 250 per spectral pixel in the middle of the helium line. From 1.8 to 1.5 h before mid-transit, the SNR plummeted to 50--100, before recovering shortly before ingress.  Due to human error, we took no data for the 15 minutes centered on 1.06 h before mid-transit, and for the 4 minutes centered on 0.66 h after mid-transit.  These gaps add up to only 11\% of the transit duration, and do not significantly affect the results.

The Keck/NIRSPEC data were reduced using the pipeline and methodology described in \cite{zhang_2022c} and \cite{zhang_2023}.  Briefly, we generate a median dark and a median flat; produce A-B difference images and divide them by the flat; use optimal extraction to extract the spectra; use a combined stellar and telluric model to obtain the wavelength solution and continuum for each spectrum; and use \texttt{molecfit} \citep{smette_2015} to correct for telluric absorption.  For our observations, there is no significant telluric absorption that overlaps with the helium line.  Telluric absorption only begins to pick up redward of 10834.5\AA{} in the stellar rest frame, corresponding to a star-relative redshift of 33 km/s.

In addition to the Keck/NIRSPEC data, we obtained a 18.2 ks XMM-Newton observation of the star on 2022-09-12 (observation ID 0903000301, PI: Michael Zhang), about three months after the helium observations.  We analyzed the EPIC X-ray data using \texttt{SAS} and fit a single-temperature thin plasma model (APEC) using XSPEC, following the same methodology as \cite{zhang_2022c}.  Simultaneously with the X-ray observations, XMM-Newton's Optical Monitor measures the mid-ultraviolet (MUV) flux in two bandpasses: UVW2 (212 nm, width 50 nm) and UVM2 (231 nm, width 48 nm).  MUV ionizes metastable helium, making the this flux an important input to models of helium absorption.

Finally, we examined publicly available photometry from the Transiting Exoplanet Survey Satellite (TESS).  \ourstar was observed in 5 sectors: 26, 40, 52, 53, and 54.  By coincidence, TESS was observing the star simultaneously with our Keck/NIRSPEC helium observations.  We see no flares or other evidence of stellar variability in the TESS data during the transit or within several hours of it, with the possible exception of a $\sim$300 ppm drop in flux 3.5 h after mid-transit.  However, this drop would have happened well after the end of our NIRSPEC observations.  Any flare 0.1\% or bigger would have been easily visible, so we rule these out with high confidence.

\section{Results}
\label{sec:results}
\subsection{Stellar age}
We originally included \planet in our sample because its rotational period and X-ray luminosity suggest an age $\lesssim$1 Gyr.  However, isochrone fitting by \cite{rescigno_2023} reveal a much older age of $3.8_{-2.7}^{+5.5}$ Gyr.

The discrepancy arises because the Lomb-Scargle periodogram of the TESS light curve for sectors 26 and 40 show a broad peak from 10--14 days, while the Second ROSAT All-Sky Survey (2RXS) reports an X-ray flux of 1\e{-13} erg s$^{-1}$ cm$^{-2}$ (power law fit) or 7\e{-14} erg s$^{-1}$ cm$^{-2}$ (blackbody fit), based on $21.2 \pm 7.5$ background-corrected counts.  Using the relations found by \cite{mamajek_2008}, the rotation period implies an age of 350--640 Myr, while the X-ray flux implies an age of 1.1--1.4 Gyr.  Since we began the survey, TESS sectors 52, 53, and 54 have become available.  Unfortunately, adding these data weakens the Lomb-Scargle peak, and even though visual inspection of the light curve reveals an unmistakable variability with a RMS of 0.15\%, it does not reveal any obvious rotation period.  \cite{rescigno_2023} likewise did not securely detect the rotation signal in WASP data spanning $\sim$850 days.

The XMM-Newton observations we obtained three months after our helium observations reveal an unexpectedly low X-ray flux of 2.3\e{-14} erg s$^{-1}$ cm$^{-2}$ in the 5--100 \AA{} bandpass, or 0.5 erg s$^{-1}$ cm$^{-2}$ at 1 AU.  This is 3--4 times lower than the flux measured by ROSAT in August 1990, suggesting that the star has become significantly quieter in the intervening 32 years, and/or that ROSAT detected an upward statistical fluctuation.  Combining this flux with the \cite{mamajek_2008} relation between age and X-ray flux, we obtain a significantly older age of 2.8 Gyr.

Finally, we estimated the age via the star's $\log{R'_{\mathrm{HK}}}$ of $-4.83 \pm 0.45$ \citep{rescigno_2023}, which translates to an age of 3.5 Gyr using the age-$\log{R'_{\mathrm{HK}}}$ relation in \cite{mamajek_2008}.  Taking the geometric mean of all estimates, we arrive at an age of $2.3 \pm 1.2$ Gyr.  The advanced age is confirmed by the low vsin(i) of $0.78 \pm 0.09$ km/s, and by the isochrone-derived age \citep{rescigno_2023}.

\subsection{Helium absorption}

\begin{figure*}[t]
  \centering 
    \includegraphics[width=0.9\textwidth]{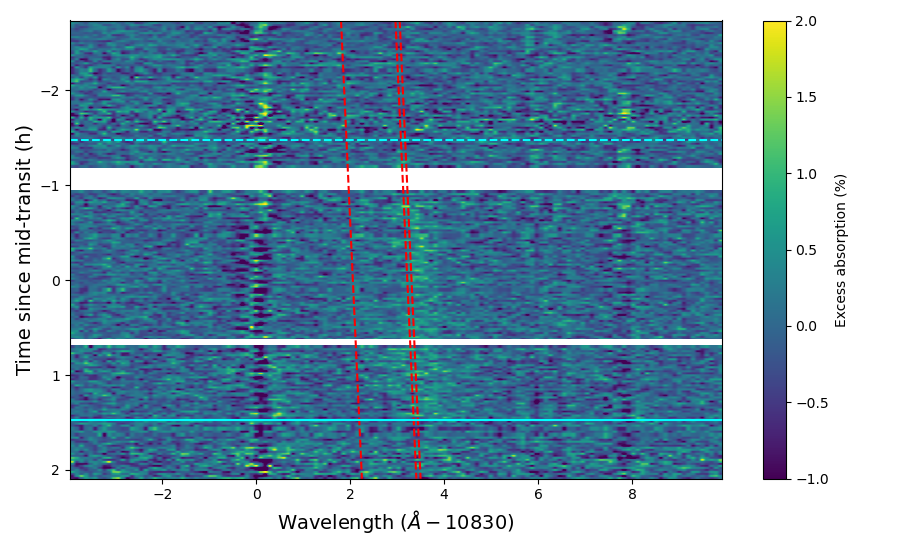}
    \caption{Percent excess absorption from \planet as a function of time and wavelength (stellar rest frame).  The dashed cyan line indicates the beginning of the white light transit, while the solid cyan line indicates the end.  The red lines show the wavelengths of planetary helium absorption.  The white bars indicate gaps in the data.  Note the variability in the stellar Si I line at 10830 \AA{} and Na I line at 10838 \AA{}, as well as the uncorrected telluric variability around 10836 \AA{}.  There is no stellar photospheric line or telluric line near the 10833 \AA{} helium absorption.}
    \label{fig:excess_2D}
\end{figure*}

\begin{figure*}
    \subfigure {\includegraphics[width=0.5\textwidth]{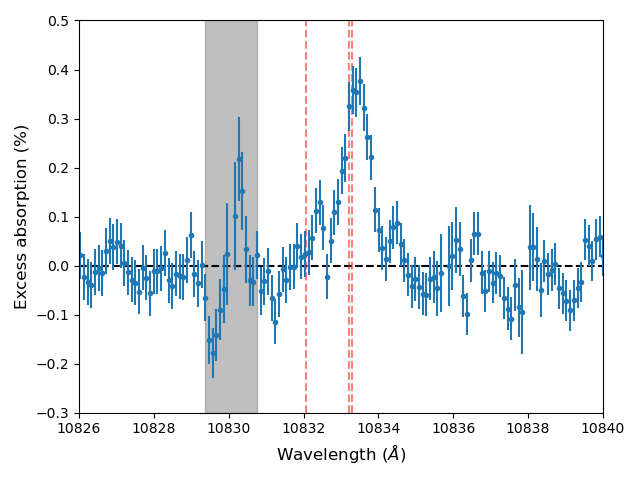}}
    \subfigure {\includegraphics[width=0.5\textwidth]{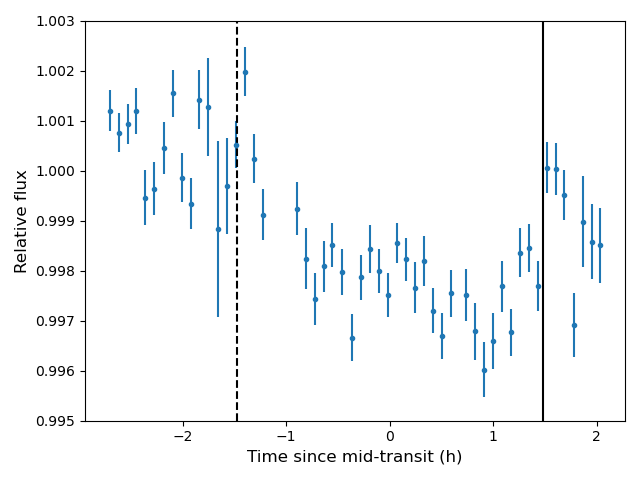}}
    \caption{Left: average in-transit excess absorption in the planet rest frame.  The deep silicon line at 10830 \AA{} poses difficulties for spectral extraction and may be inherently variable, causing anomalies in the 0.6 \AA{} surrounding that wavelength (gray region).  Right: light curve of the helium line, defined as the region within 0.75 \AA{} of 10833.3 \AA{} in the stellar rest frame.  This bandpass only captures the stronger peak.  The two vertical lines mark the beginning and end of white light transit.}
    \label{fig:absorption_1D}
\end{figure*}

The excess absorption as a function of wavelength and time is shown in Figure \ref{fig:excess_2D}.  In previous papers, we masked strong stellar and telluric lines in plots to avoid confusing the reader, but here we leave all wavelengths unmasked in order to show the variability in these lines.  The variability around 10836 \AA{} is due to the one and only strong telluric absorption region within this bandpass.  \texttt{molecfit} mostly succeeds in correcting the 8\% absorption, but leaves behind residuals of $\sim$0.5\%.  There are two strong photospheric lines: Si I 10830 \AA{} and Na I 10838 \AA{}.  Both stellar lines show an abrupt increase in flux around 0.65 h before mid-transit, but the helium line appears unaffected.  This increase is uncorrelated with the planet--it occurs well after ingress, when helium absorption is already evident, and unlike the helium absorption, it does not reverse at egress.  This percent-level variability in equivalent width was previously seen for TOI 1430b, but not for 560b or 2076b (1683b's SNR was too low for a proper comparison); however, variability in the Si line shape was also seen for 560b and 2076b, and variability in the Na line shape was visually evident for 560b.  We suspect the variability may be related to changes in the instrumental line spread profile, but cannot rule out stellar variability.  Although we have not seen increased systematics in the helium line that correlate with variability in the Si and Na lines, further work will be necessary to explain the cause of the variability and estimate its potential impact on the helium results.

Figure \ref{fig:absorption_1D} (left) shows the average in-transit excess absorption spectrum.  We integrate the part of the excess absorption spectrum between 10831 and 10835 \AA{} to obtain the equivalent width, $W_{\rm avg}=3.3 \pm 0.3$ m\AA{}.
In the absorption spectrum, the peak value is $0.37 \pm 0.05$\% and occurs at a redshift of $7 \pm 3$ km/s.  This redshift can also be seen in the 2D excess absorption plot (Figure \ref{fig:excess_2D}).  Radial velocity data \citep{rescigno_2023} shows that the planet's radial velocity at mid-transit is -$Ke\cos{\omega} = 1.3_{-1.3}^{+2.5}$ km/s, consistent with the measured redshift to 1.5$\sigma$.  The absorption spectrum also shows a secondary peak (from the third line of the helium triplet) with a $10 \pm 3$ km/s redshift and a peak absorption of $0.13 \pm 0.05$\%.  We consider the secondary peak detection to be likely but not conclusive because there are $\sim0.05$\% peaks and valleys in the spectrum due to correlated noise, including a $0.09$\% bump redward of the main helium peak.  If the secondary peak is real, the ratio between the two peaks would be $0.35 \pm 0.13$--in between the optically thin limit of 0.125 and the perfectly optically thick limit of 1.  For comparison, of the four planets in \cite{zhang_2023}, only TOI 2076b had a peak ratio inconsistent with an optically thin outflow.  If the outflow is not entirely optically thin, the secondary peak would trace gas closer to the planet, which could explain part of the (statistically insignificant) difference in redshift between the two peaks.  

Figure \ref{fig:absorption_1D} (right) shows the light curve of a 1.5 \AA{} region centered on the helium line.  The pre-transit flux appears to fluctuate more than the theoretical error bars would suggest, probably due to a combination of systematics, weather, and stellar activity, but shows no trend.  The flux drops after ingress, but does not reach its minimum until almost 1 h after mid-transit.  After egress, the flux quickly recovers--though not quite to its pre-transit value, which could be due to either stellar activity or a long tail of outflowing gas.  The light curve asymmetry is also reflected in the 2D excess absorption plot.  For example, the region between ingress and the data gap shows less absorption than the equivalent region right before egress.

Stellar variability in the helium line is poorly understood.  To help assess the extent to which stellar variability might have affected our observations, we collected 1.7 h of out-of-transit monitoring data on July 4, 2023.  This data shows limited stellar variability, with the band-integrated light curve exhibiting a standard deviation of 0.063\% and no secular trend (Appendix \ref{section:oot_monitoring}).


\begin{table}[ht]
    \caption{Outflow properties}
    \centering
    \setlength{\tabcolsep}{6pt}
    \begin{tabular}{c c}
        \hline
	  Peak absorption & $0.38 \pm 0.05$\% \\
        Redshift*    & $6.7 \pm 2.7$ km/s\\
        Peak ratio  & $0.34 \pm 0.13$\\
        W\textsubscript{avg} & $3.3 \pm 0.3$ m\AA{}\\
        $\dot{m}_{\rm OOM}$ & 5.5\e{9} g/s\\ 
        $\dot{m}_{\rm Parker}$ & $5.5 \pm 0.9$ \e{9} g/s\\
        $T_{\rm Parker}$ & $4640 \pm 230$ K\\
        Redshift$_{\rm Parker}$* & $4.7 \pm 0.5$ km/s\\
	\hline 
    \end{tabular}
    \tablenotetext{*}{$1.3_{-1.3}^{+2.4}$ km/s of this redshift is due to the planet's eccentricity}
    \label{table:outflow_properties}
\end{table}

To estimate the mass loss rate implied by our observations, we use the same two methods as \cite{zhang_2023}: an order-of-magnitude (OOM) method, and a Parker wind method that uses the 1D spherically symmetric isothermal model of \cite{oklopcic_2019}.  We do not expect either method to be accurate to more than a factor of a few.  The order-of-magnitude method assumes the outflow is optically thin, that the outflow speed is always the sound speed $c_s$, and that $f=10^{-6}$ of the helium atoms are in the metastable ground state.  Under these assumptions, the mass loss rate can be derived from the equivalent width of the helium absorption.  (If the outflow is not optically thin, the mass loss rate would be underestimated, potentially by a factor of a few.)  To recap the derivation in \cite{zhang_2023}, the equivalent width gives the number of metastable helium atoms currently in front of the star; dividing by the replacement timescale (roughly, the stellar radius divided by the outflow speed) gives the mass loss rate of metastable helium, while further dividing by the mass fraction of metastable helium gives the total mass loss rate of the whole outflow.  The result is:

\begin{align} \label{eq:obs_rate}
     \dot{m}_{\rm obs} = \frac{R_* m_e m_{He} c_s c^2 W_{\rm avg}}{0.25f e^2 \lambda_0^2 \sum{g_l f_l}},
\end{align}

where $W_{\rm avg}$ is the equivalent width, we assume $c_s$=10 km/s, and the 0.25 comes from the assumption that 25\% of the mass of the outflow is in helium atoms or ions.  $\sum{g_l f_l}$, the sum of the product of the degeneracy and oscillator strength, is either 1.44 (if summed over the two inseparable lines) or 1.62 (if summed over all three lines).  We adopt 1.62, but the 12\% difference is negligible compared to the uncertainties in the other quantities.

The Parker wind method requires a stellar spectrum.  We construct one using the same methodology as in \cite{zhang_2023}, obtaining a spectrum (Figure \ref{fig:stellar_spectrum}) with $F_X=0.45$, $F_{\rm EUV_{He}}=2.4$ (100-504 \AA{}), $F_{\rm EUV}=3.5$ (100--912 \AA{}), and $F_{\rm MUV}=7.8$ (1230--2588 \AA{}), all reported at 1 AU in units of erg s$^{-1}$ cm$^{-2}$.  The combined X-ray and EUV flux at the planet, 650 erg s$^{-1}$ cm$^{-2}$, is several times lower than the 5000--12,000 experienced by the four young mini-Neptunes in \cite{zhang_2023}.  To our knowledge, it is the lowest XUV flux of any planet with a escaping helium detection (see catalog in Appendix), the second lowest being HD 209458b's 1000 erg s$^{-1}$ cm$^{-2}$ \citep{alonso-floriano_2019}.  It is important to note that all EUV fluxes are reconstructed because no current telescope can observe even the EUV from the nearest stars, and the EUV reconstruction is uncertain by an order of magnitude \citep{france_2022}.  
After obtaining the stellar spectrum, we run the Parker wind model of \cite{oklopcic_2019} in a nested sampling framework with the following parameters: the log of the mass loss rate, the temperature, and a blueshift.  Table \ref{table:outflow_properties} shows the inferred parameters.  As with the planets reported in \cite{zhang_2023}, the width of the line implies an outflow with a temperature of several thousand K, consistent with photoevaporation but not with core-powered mass loss.  This conclusion assumes that no mechanism significantly broadens the absorption beyond the width predicted by a Parker wind.

Largely coincidentally, the OOM method and the Parker wind method both estimate a mass loss rate of 5.5\e{9} g/s.  Assuming an envelope fraction of 1\% and no change in the mass loss rate, the envelope lifetime would be 3.1 Gyr.  The decrease in stellar high-energy flux with age and the shrinking of the planet as its envelope is stripped should both decrease the mass loss rate in the future, while increasing it in the past.  Nevertheless, it is reassuring that as with the other four planets, the lifetimes we infer are comparable to the age of the planet.

The mass loss rate can be compared with the energy-limited mass loss rate, a theoretical maximum which assumes all the energy from the incoming XUV flux goes into lifting gas out of the gravity well:

\begin{align} \label{eq:energy_lim_rate}
    \dot{m}_{\rm theory} &= \pi \frac{R_{\rm XUV}^3 F_{\rm XUV}}{G M_p}\\
    &= \frac{3F_{\rm XUV}}{4G \rho_{\rm XUV}}
\end{align}

$R_{\rm XUV}$ is the XUV photosphere radius.  To roughly estimate this radius, we slightly follow \citet{wang_2018} in assuming that the EUV photosphere is at $\rho=10^{-13}$ g cm\textsuperscript{-3} and that the atmosphere is isothermal between the optical photosphere and the XUV photosphere.  With these assumptions, we calculate $R_{\rm EUV} = 1.25 R_p$ and $\dot{m} = 3.9 \times 10^9$ g/s.  This theoretical maximum is $\sim$2x smaller than the order-of-magnitude ``observed'' mass loss rate.  Given the very large error bars on all quantities, we consider the observations consistent with an efficient outflow.  A highly efficient outflow is expected for a planet of such low gravitational potential and low XUV flux.  For example, \cite{caldiroli_2022} ran a suite of 1D hydrodynamic simulations using their ATES code and found an efficiency of 96\% for a hypothetical Neptune-like planet with similar XUV irradiation and gravitational potential as \planet (their Figure 2).

\section{Discussion}
\label{sec:discussion}

We have two equations for the mass loss rate: the ``order-of-magnitude'' expression proportional to equivalent width (Equation \ref{eq:obs_rate}) and the energy-limited formula (Equation \ref{eq:energy_lim_rate}).  The two should therefore be equal.  If we drop the constants and allow for some dependence of $\eta$ on $F_{\rm XUV}$ and $\rho_{\rm XUV}$, we can make the weaker prediction that $R_* W_{\rm avg}$ must be positively correlated with $F_{\rm XUV}/\rho_{\rm XUV}$ among planets with helium detections.  We note that \cite{vissapragada_2022} plotted the mass loss rate inferred from the Parker wind model against $F_{\rm XUV}/\rho$ for their 7 helium survey targets, but the large error bars and limited sample size prevented the detection of any trend.

\begin{figure*}[ht]
    \includegraphics[width=\textwidth]{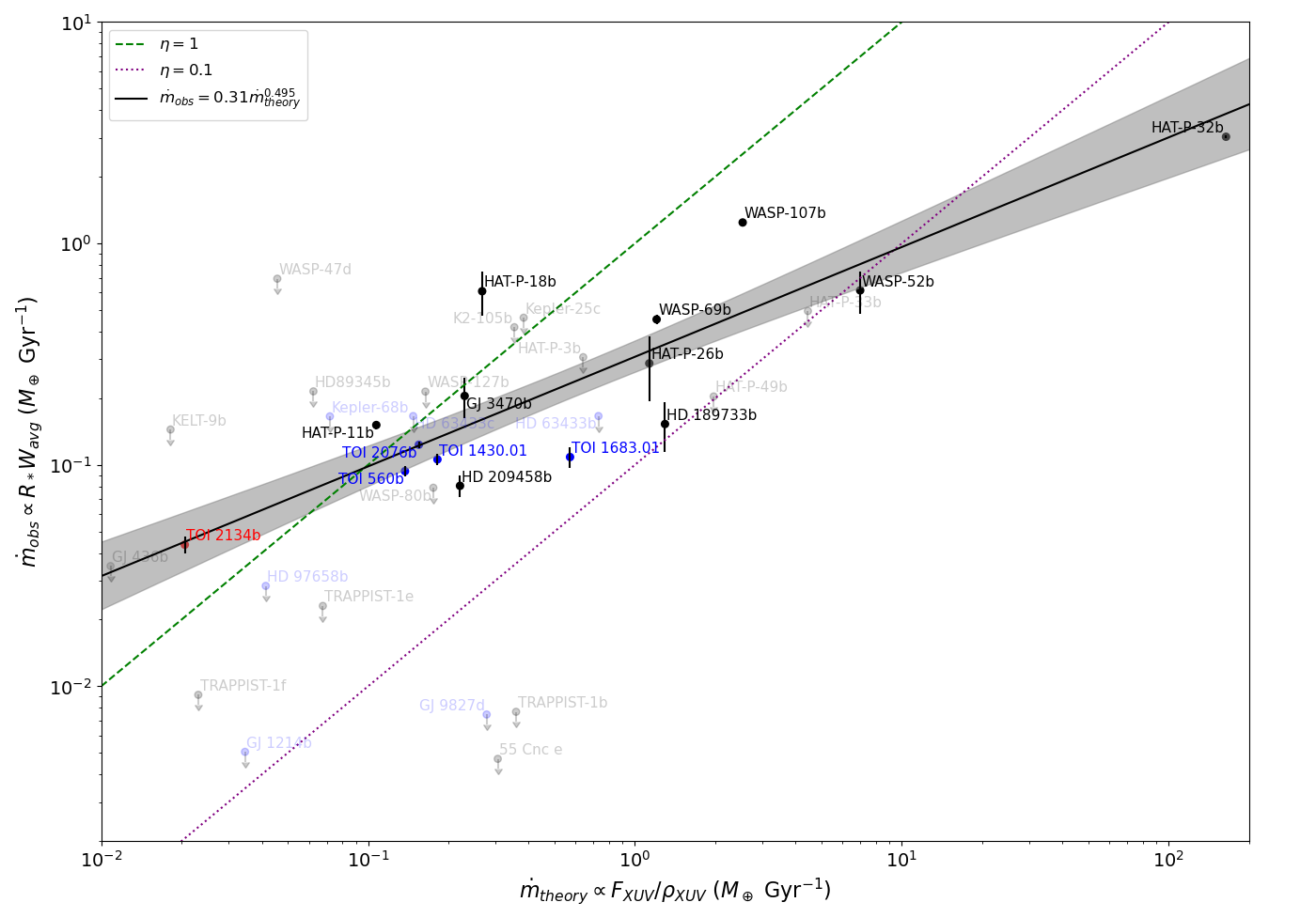}
    \caption{Relationship between the ``observed'' and ``energy-limited'' mass loss rates, with mini-Neptunes highlighted in red (\planet) or blue (all others).  Stripping off the constants, this is a relationship between $R_* W_{\rm avg}$ and $F_{\rm XUV}/\rho_{\rm XUV}$.  For consistency with previous literature, we define XUV to be 5--504 \AA{}.  The dashed lines indicate what the relationship should be for 100\% efficiency and 10\% efficiency.  Omitted for clarity is the very large (potentially order-of-magnitude) uncertainty on the XUV flux, and therefore on the x axis values, for all data points.}
    \label{fig:planets_context}
\end{figure*}

To test our prediction, we gathered $F_{\rm XUV}/\rho_{\rm XUV}$ and $R_* W_{\rm avg}$ for all published helium detections.  As we describe in the Appendix, this was challenging to do in an accurate and consistent way, but we made an effort to maximize consistency without reanalyzing each detection.  Figure \ref{fig:planets_context} plots the data we collected, and shows a strong positive correlation between $\log(\dot{m}_{\rm theory})$ and $\log(\dot{m}_{\rm obs})$.  We fit a power law to the data ($\dot{m}_{\rm obs} = \eta_0 (\dot{m}_{\rm theory})^P$) by taking the log of both sides and using scipy's Orthogonal Distance Regression (ODR), which takes into account errors in both the independent and dependent variables.  ODR results do not change if all errors are inflated or deflated by the same factor.  We assume equal errors in log space for all data points and both variables--a concession to the fact that the XUV flux, mass loss efficiency, and conversion factor between equivalent width and mass loss rate all have large uncertainties of at least a factor of a few, and that these uncertainties swamp the observational error.  With these approximations, we obtain $\eta_0 = 0.30 \pm 0.06$ and $P = 0.50 \pm 0.08$;  applying Student's t-test, we find that a zero slope is ruled out with p=2.3\e{-5}.  Using the Spearman test, which only tests for monotonicity and does not assume a linear relationship, the trend remains significant (p=1.7\e{-4}).  The trend also remains significant even after removing the two extreme points, \planet and HAT-P-32b (ODR p=0.0036, Spearman p=0.005).

Another notable fact about Figure \ref{fig:planets_context} is that the efficiency appears to decrease with increasing $F_{\rm XUV}/\rho_{\rm XUV}$.  This would not be surprising, as many previous works have found that efficiency is lower at high irradiation levels (e.g. \citealt{caldiroli_2022,zhang_2022c}), a consequence of most of the energy being radiated away by recombination in this ``recombination-limited'' regime \citep{lampon_2021}.  However, the sub-linearity of the relation ($p < 1$) could also be an artifact of our assumption that $10^{-6}$ of all helium atoms are in the triplet state.  HAT-P-32b, a planet orbiting a relatively hot star ($6207 \pm 88$ K; \citealt{bonomo_2017}), should have a much lower triplet helium fraction than the K dwarf planet \planet.  We attempt a crude correction for the dependence of the triplet helium fraction on stellar temperature and semimajor axis, finding that even though it does result in a linear relationship between $R_* W_{avg}$ and $F_{XUV} / \rho_{XUV}$, it also weakens the correlation.  In the Appendix, we describe this variant of the correlation in more detail, in addition to describing attempts to account for the star's gravitational potential, differing mass loss efficiencies, and different estimates of planet density.  We find that the relation is statistically significant regardless of these choices.

Finally, it is informative to examine the non-detections in Figure \ref{fig:planets_context}. We excluded detections that were claimed to be tentative (notably the tentative GJ 1214b detection of \citealt{orell-miquel_2022}), and included only the most sensitive non-detection where multiple exist.  For spectroscopic non-detections where only an upper limit on the percent excess absorption is reported, we multiply by an effective width of 1 \AA{} (a typical value for the spectroscopic detections) to obtain an upper limit on the equivalent width.  Non-detections of planets smaller than 2 $R_\Earth$ (55 Cnc e, TRAPPIST-1 b/e/f) are unsurprising as these planets likely have no H/He atmosphere.  Most non-detections are unconstraining, as they fall above the trend-line (e.g. the mini Neptunes Kepler-68b and HD 63433c).  The non-detections of HD 63433b and HD 97658b are weakly constraining, and multiple papers have been written on the non-detection of WASP-80b \citep{fossati_2022,vissapragada_2022,fossati_2023}.  The high XUV flux calculated by \cite{fossati_2022} would make the non-detection surprising, while the low XUV flux calculated by \cite{fossati_2023} would not.  However, some non-detections seem highly constraining, including the \cite{kasper_2020} non-detection of GJ 1214b (also see the non-detection of \citealt{spake_2022}), which falls 1.1 dex below the trend line, and that of GJ 9827d, which falls 0.5 dex below even the $\eta=0.1$ line.  Recent JWST/MIRI observations of GJ 1214b suggest its atmosphere is of high mean molecular weight \citep{kempton_2023}; however, \cite{orell-miquel_2022} report a tentative detection of escaping helium.  GJ 9827d is a small planet (2.0 $R_\Earth$) orbiting a several Gyr old star \citep{rice_2019}, and \cite{carleo_2021} suggests that it may have lost any H/He envelope, although its low density of 2.5 g cm$^{-3}$ would be puzzling if that were the case.  We encourage further helium observations of mini-Neptunes around mature stars to determine which ones have helium outflows and which ones do not.

\section{Conclusion}
\label{sec:conclusion}
In this paper, we presented the fifth detection of escaping helium from a mini-Neptune, and the first definitive detection from a mature mini-Neptune.  Among all helium detections, it has the lowest equivalent width, and comes from the planet receiving the lowest XUV flux.  The width of the helium signal implies a photoevaporative origin, while the equivalent width implies a mass loss timescale in the Gyr range.

Putting \planet in the context of other helium detections, we observe the theoretically expected positive correlation between $F_{\rm XUV}/\rho_{\rm XUV}$ and $R_* W_{\rm avg}$ to high statistical significance.  \planet, which has the lowest value along both axes, anchors the lower left side of the relation.  This relation demonstrates that currently published helium measurements are sufficient to detect statistical patterns, and may be sufficient to test mass loss simulations at the population level.  We encourage further observations to fill in the 1.4 dex gap in $F_{\rm XUV}/\rho_{\rm XUV}$ between WASP-52b and HAT-P-32b at the high end, and if possible, to fill in the smaller gap between \planet and HAT-P-11b at the low end.  We also encourage researchers reporting new helium detections to report the equivalent width in addition to the peak excess absorption, because the EW is both more directly correlated with the mass loss rate and less sensitive to differing instrumental resolutions.  Finally, as is well known, the stellar XUV flux is highly uncertain.  We encourage further efforts to characterize the high-energy output of these stars, whether by measuring the X-ray spectrum (e.g. \citealt{foster_2022}), measuring  Ly$\alpha$ and metal lines in the FUV (e.g.. \citealt{bourrier_2018}), or launching a space telescope that can obtain direct measurements of the EUV flux \citep{france_2022}.

\textit{Software:}  \texttt{numpy \citep{van_der_walt_2011}, scipy \citep{virtanen_2020}, matplotlib \citep{hunter_2007}, dynesty \citep{speagle_2019}, SAS \citep{gabriel_2004}, XSPEC \citep{arnaud_1996}, \citep{smette_2015}}

\section{Acknowledgments}
We thank Dakotah Tyler for his help in collecting the observations.  We thank Jaume Orell-Miquell and colleagues for helpful discussions about 1D modelling.  

The helium data presented herein were obtained at the W. M. Keck Observatory, which is operated as a scientific partnership among the California Institute of Technology, the University of California and the National Aeronautics and Space Administration. The Observatory was made possible by the generous financial support of the W. M. Keck Foundation.  

We used observations obtained with XMM-Newton (observation ID 0903000301), an ESA science mission with instruments and contributions directly funded by ESA Member States and NASA.  We acknowledge funding from XMM-Newton grant 80NSSC22K0742.  

Funding for the TESS mission is provided by NASA's Science Mission Directorate.  We acknowledge the use of public TESS data from pipelines at the TESS Science Office and at the TESS Science Processing Operations Center.

MZ acknowledges support from the 51 Pegasi b Fellowship funded by the Heising-Simons Foundation.

FR is funded by the University of Exeter's College of Engineering, Maths and Physical Sciences, UK.

\appendix

\section{Telluric correction}
\begin{figure*}[ht]
\centering
    \includegraphics[width=0.6\textwidth]{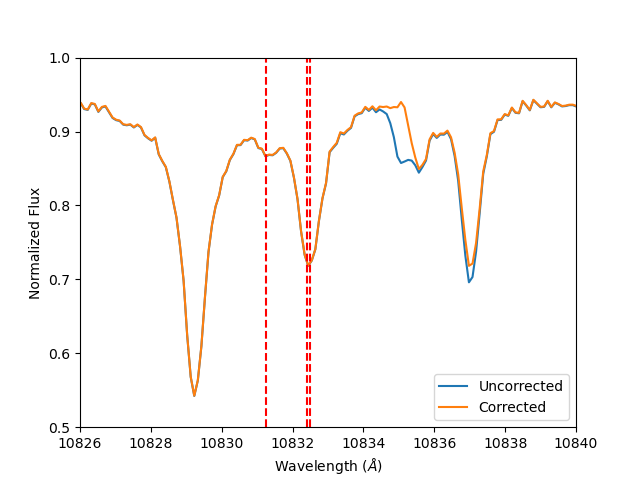}
    \caption{The stellar spectrum in the terrestrial frame, before and after correction for telluric absorption.}
    \label{fig:telluric_correction}
\end{figure*}

Figure \ref{fig:telluric_correction} shows the impact of telluric correction by \texttt{molecfit}.  As can be seen, there are no strong telluric absorption lines near the helium line.  There is a weak telluric emission line overlapping the helium line with an amplitude of $\sim$20 electrons/pixel, compared to $\sim$11,000 electrons/pixel on the trace.  Emission lines are not a concern because they are subtracted out very effectively by the ABBA nod pattern.

\section{High-energy spectrum}

\begin{figure*}[ht]
\centering
    \includegraphics[width=0.5\textwidth]{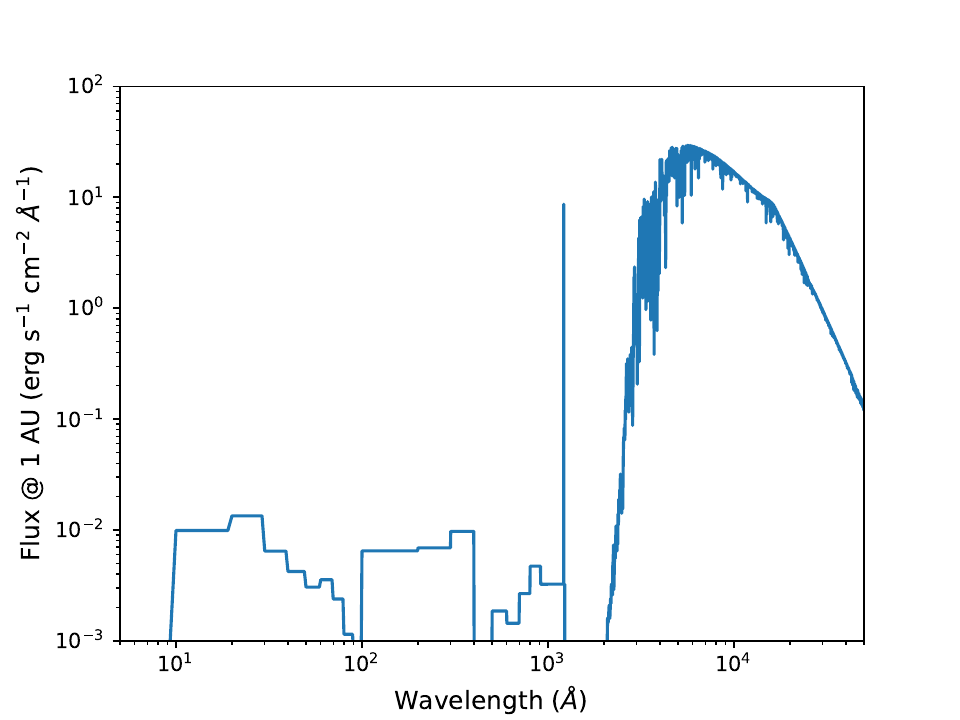}
    \caption{The reconstructed stellar spectrum.}
    \label{fig:stellar_spectrum}
\end{figure*}

Figure \ref{fig:stellar_spectrum} shows the reconstructed stellar spectrum.  The X-ray spectrum is derived from XMM-Newton observations, and the MUV flux is consistent with XMM-Newton photometric observations to within 13\%.  Ly$\alpha$ and EUV are reconstructed from the X-ray luminosity.

As in \cite{zhang_2023}, we used XSPEC to fit a thin plasma model (APEC) to the X-ray observations, obtaining $kT=0.219 \pm 0.013$ keV, EM=$0.44 \pm 0.05 \times 10^{50}$ cm$^{-3}$, and $F_X = 2.1 \pm 0.3 \times 10^{-14}$ erg s$^{-1}$ cm$^{-2}$ (0.124--2.48 keV).  Note that EM is roughly inversely proportional to metallicity, and because we do not fit the metallicity, the EM value should only be trusted if the coronal metallicity is solar.

There are a few differences between our analysis in this paper and in \cite{zhang_2023}.  First, for unknown reasons, the pn detector spectrum that SAS generates with default settings is nearly zero.  We therefore reran SAS after manually defining source and background regions for all three detectors: the source region is a circle of radius 20 arcsec centered on the star's expected coordinates (computed with Gaia DR3 data), while the background region is an annulus centered on the same point, with an inner radius of 30 arcsec and an outer radius of 60 arcsec.  Second, we fix the coronal metallicity to solar instead of fitting it because there were not enough photons to give a good constraint.  Third, we take into account interstellar photoelectric absorption, with a fixed hydrogen column density of $10^{18.25}$ g/cm$^2$ (near the upper end for stars at similar distances; \citealt{wood_2005}).  Including interstellar absorption increased the inferred flux by less than a few percent.

\section{Out of transit activity monitoring}
\label{section:oot_monitoring}
On July 4, 2023, when TOI 2134b was days away from the nearest transit, we observed TOI 2134 for 1.7 h using Keck/NIRSPEC to monitor the out-of-transit variability of the stellar helium line.  We used exactly the same settings, and analyzed the data in exactly the same way, as for the science observations.  In the middle of the observations, a telescope fault occurred, causing a 12 minute gap in the data.

\begin{figure*}[ht]
  \centering 
    \includegraphics[width=0.9\textwidth]{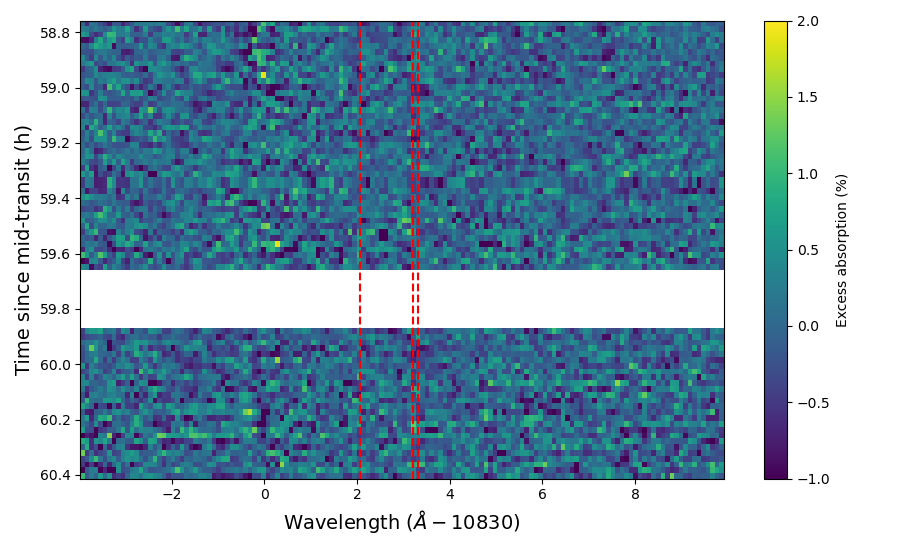}
    \caption{Percent excess absorption (relative to median) as a function of time and wavelength, while the planet was days away from transit.  This plot is of the same format, and has the same wavelength range and colorbar scale, as \ref{fig:excess_2D}.}
    \label{fig:excess_2D_no_transit}
\end{figure*}

\begin{figure*}[ht]
  \centering 
    \includegraphics[width=0.5\textwidth]{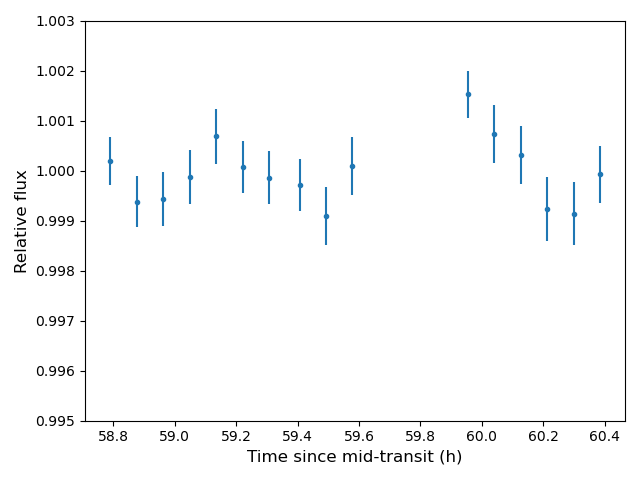}
    \caption{Light curve of the helium line (within a half-width of 0.75 \AA{}) while the planet was far from transit.  This plot is the analogue of Figure \ref{fig:absorption_1D} (right), and the y axis has the same scale.}
    \label{fig:light_curve_toi2134_notransit}
\end{figure*}

Figure \ref{fig:excess_2D_no_transit} shows the excess absorption (relative to the median) as a function of time and wavelength, while Figure \ref{fig:light_curve_toi2134_notransit} shows the light curve of the helium line.  The helium line shows no signs of variability in the 2D excess absorption plot, although the Si I 10830 \AA{} line is again variable at the 1\% level.  The light curve exhibits variability of 0.063\%--slightly higher than the typical photon error of 0.054\%, but far lower than the 0.3\% helium absorption observed during the science observations.  The monitoring observations are consistent intrinsic stellar variability of 0--0.08\%.

\section{Parker wind fit}

\begin{figure*}[ht]
  \includegraphics[width=\textwidth]{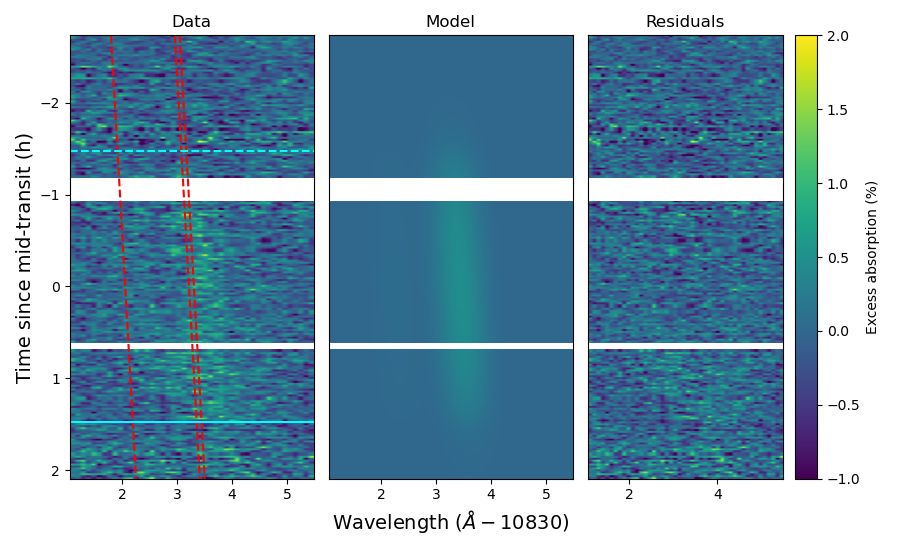}
    \caption{The best fit from our Parker wind model, compared to the data and the residuals.}
    \label{fig:model_fit}
\end{figure*}

\begin{figure*}[ht]
\centering
  \includegraphics[width=0.8\textwidth]{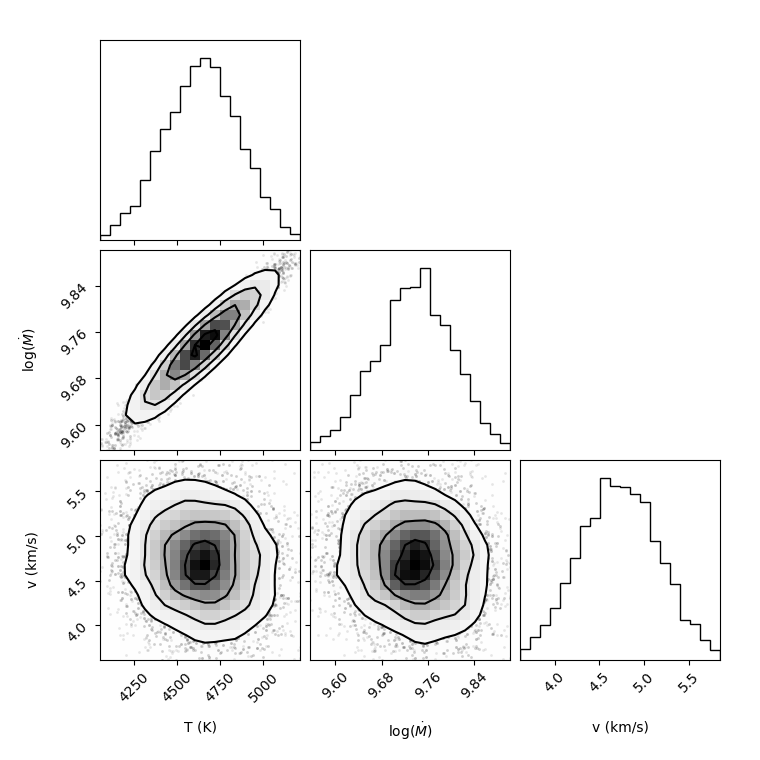}
    \caption{2D posteriors from nested sampling for the Parker wind model.}
    \label{fig:corner}
\end{figure*}

In Figure \ref{fig:model_fit}, we show the best fit from the Parker wind model; in Figure \ref{fig:corner}, we show the posterior distribution.

In \cite{zhang_2023}, we imposed a cutoff of one Hill radius on the outflow, for the reason that the spherically symmetric model may break down beyond that point.  This meant that we ignored all helium absorption originating from outside the cutoff.  However, gas that flows beyond the cutoff does not disappear, and may not even become significantly unspherical (see e.g. the 3D simulations of \citealt{khodachenko_2019,zhang_2022a}).  For this paper, we therefore choose a cutoff radius of 1 $R_*$, much larger than the Hill sphere of 0.54 $R_*$.  Had we chosen the Hill sphere as the cutoff radius, the inferred mass loss rate would have been $1.1 \times 10^{10}$ g/s, 2.2x our fiducial value.

\section{The trend}

Table \ref{table:helium_detections} shows the helium detections and non-detections plotted in Figure \ref{fig:planets_context}.

\begin{table}[ht]
    \caption{Helium detections and non-detections.  Tentative detections are excluded, as are targets with no reported XUV flux, and giant planets without mass measurements.  All $F_{\rm XUV}$ estimates have uncertainties of at least a factor of a few.}
    \centering
    \setlength{\tabcolsep}{6pt}
    \begin{tabular}{c C C C C C c C c}
        \hline
        Planet & F_{\rm XUV} (5-504 \text{\AA}) & R_p & \rho_{\rm XUV}    & W_{\rm avg}   & \sigma_W & Detected & R_* & Ref\\
               & (10^3 \; \rm erg \; s^{-1} cm^{-2}) & (R_\Earth) & ({\rm g/cm}^3) & \text{(m\AA{})} &  \text{(m\AA{})} &  & ($R_\Sun$) \\
	\hline
        WASP-69b & 4.17 & 11.85 & 0.21 & 28.5 & 1.5 & True & 0.86 & \cite{nortmann_2018} \\
        HD 189733b & 19.2 & 12.54 & 0.88 & 11.0 & 2.8 & True & 0.75 & \cite{zhang_2022c} \\
        HD 209458b & 1.004 & 15.23 & 0.27 & 3.65 & 0.4 & True & 1.19 & \cite{alonso-floriano_2019} \\
        HAT-P-11b & 2.109 & 4.36 & 1.17 & 12.0 & 0.56 & True & 0.68 & \cite{allart_2018} \\
        WASP-107b & 2.664 & 10.4 & 0.06 & 100.0 & 3.3 & True & 0.67 & \cite{kirk_2020} \\
        GJ 3470b & 1.435 & 4.57 & 0.37 & 20.1 & 4.1 & True & 0.55 & \cite{palle_2020} \\
        GJ 1214b & 0.64 & 2.742 & 1.10 & 1.3 & 0.79 & False & 0.21 & \cite{kasper_2020} \\
        HAT-P-32b & 162.0 & 22.19 & 0.06 & 118.4 & 7.1 & True & 1.37 & \cite{czesla_2022} \\
        WASP-52b & 25.0 & 13.71 & 0.21 & 42.0 & 9.0 & True & 0.79 & \cite{kirk_2022} \\
        GJ 436b & 0.197 & 4.191 & 1.08 & 4.1 & 2.05 & False & 0.46 & \cite{nortmann_2018} \\
        KELT-9b & 0.15 & 19.99 & 0.49 & 3.3 & 1.65 & False & 2.36 & \cite{nortmann_2018} \\
        WASP-127b & 0.058 & 14.69 & 0.02 & 8.7 & 3.6 & False & 1.33 & \cite{dos_santos_2020} \\
        GJ 9827d & 2.45 & 2.022 & 0.52 & 0.67 & 0.41 & False & 0.602 & \cite{kasper_2020} \\
        HD 97658b & 1.1 & 2.4 & 1.58 & 2.1 & 1.3 & False & 0.73 & \cite{kasper_2020} \\
        55 Cnc e & 5.8 & 1.9 & 1.13 & 0.27 & 0.16 & False & 0.94 & \cite{zhang_2021} \\
        HAT-P-18b & 0.7 & 11.1 & 0.16 & 44.0 & 10.0 & True & 0.749 & \cite{vissapragada_2022} \\
        HAT-P-26b & 2.4 & 6.3 & 0.13 & 19.7 & 6.4 & True & 0.788 & \cite{vissapragada_2022} \\
        HD 63433b & 10.3 & 2.08 & 0.84 & 10.0 & 2.0 & False & 0.897 & \cite{zhang_2022a} \\
        HD 63433c & 2.5 & 2.57 & 1.01 & 10.0 & 2.0 & False & 0.897 & \cite{zhang_2022a} \\
        TRAPPIST-1b & 9.6 & 1.116 & 1.59 & 3.467 & 1.7335 & False & 0.1192 & \cite{krishnamurthy_2021} \\
        TRAPPIST-1e & 1.5 & 0.92 & 1.32 & 10.458 & 5.229 & False & 0.1192 & \cite{krishnamurthy_2021} \\
        TRAPPIST-1f & 0.87 & 1.045 & 2.24 & 4.143 & 2.0715 & False & 0.1192 & \cite{krishnamurthy_2021}\\
        WASP-80b & 1.721 & 11.1 & 0.58 & 7.0 & 3.5 & False & 0.61 & \cite{fossati_2022} \\
        HAT-P-3b & 7.968 & 10.5 & 0.74 & 19.0 & 6.3 & False & 0.87 & \cite{guilluy_2023} \\
        HAT-P-33b & 6.195 & 20.7 & 0.08 & 14.0 & 4.7 & False & 1.91 & Ibid \\
        HAT-P-49b & 14.51 & 17.8 & 0.44 & 6.0 & 2.0 & False & 1.833 & Ibid \\
        HD89345b & 0.244 & 7.4 & 0.23 & 7.0 & 2.3 & False & 1.657 & Ibid \\
        K2-105b & 14.69 & 3.59 & 2.48 & 23.3 & 7.8 & False & 0.97 & Ibid \\
        Kepler-25c & 1.019 & 5.217 & 0.16 & 18.6 & 6.2 & False & 1.34 & Ibid \\
        Kepler-68b & 1.176 & 2.357 & 0.97 & 7.2 & 2.4 & False & 1.243 & Ibid \\
        WASP-47d & 0.577 & 3.567 & 0.75 & 32.9 & 11.0 & False & 1.137 & Ibid \\
        TOI 560b & 3.1 & 2.79 & 1.34 & 7.76 & 0.44 & True & 0.65 & \cite{zhang_2022c} \\

        TOI 1430b & 4.3 & 2.2 & 1.41 & 7.3 & 0.4 & True & 0.784 & Ibid, \cite{orell_miquel_2023}\\
        TOI 1683.01 & 7.4 & 2.6 & 0.77 & 9.22 & 0.98 & True & 0.636 & Ibid \\
        TOI 2076b & 6.7 & 2.52 & 1.34 & 8.73 & 0.3 & True & 0.762 & Ibid \\
        \planet & 0.46 & 2.69 & 1.33 & 3.32 & 0.29 & True & 0.709 & This work\\
        \hline
    \end{tabular}  
    \label{table:helium_detections}
\end{table}

\subsection{The difficulties of collating helium detections}
Creating Table \ref{table:helium_detections} in a consistent and accurate way was not easy.  Although X-ray measurements exist for many of the stars, EUV measurements do not because no EUV observatory exists, resulting in order-of-magnitude uncertainties in EUV flux and large discrepancies between different EUV estimation techniques (c.f. \citealt{france_2022}).  In addition, some authors define EUV as 5--912 \AA{} (with the hydrogen ionization energy as the upper limit), but most have chosen to define EUV as 5--504 \AA{} (with the helium first ionization energy as the upper limit).  Fortunately, most papers that have followed the former convention are ours, so we succeeded in calculating the 5--504 \AA{} flux for all stars and adopted it as the XUV flux.  The equivalent widths were even more difficult to obtain.  For the very few papers that report the equivalent width directly (e.g. HAT-P-32b; \citealt{czesla_2022}), we use the reported value.  For photometric measurements (e.g. HAT-P-26b and HAT-P-18b; \citealt{vissapragada_2022}), we obtain the equivalent width by multiplying the excess absorption by the FWHM bandwidth.  For spectroscopic measurements where the authors report no equivalent width, we integrate the excess absorption spectrum over wavelength, using the fitted model as the spectrum if one is provided (e.g. WASP-69b; \citealt{nortmann_2018}), or using the data points directly if no model is provided.  For the detections we published (HD 189733b, TOI 560b/1430b/1683.01/2076b/2134b), we integrate the data points in the excess absorption spectrum from 10831 to 10835 \AA{}.  Note that this is not how we calculated the equivalent width in \cite{zhang_2022c}, in which we used the bottom of a 1.5\AA{} bandpass, but we adopt this method for consistency (no other authors report their light curves in the exact same bandpass).  When multiple measurements are reported for the same target, we adopt the one with higher SNR if they are consistent, or take an average if they are inconsistent.

\subsection{Other variants of the correlation}
\begin{figure*}[ht]
\centering
  \includegraphics[width=\textwidth]{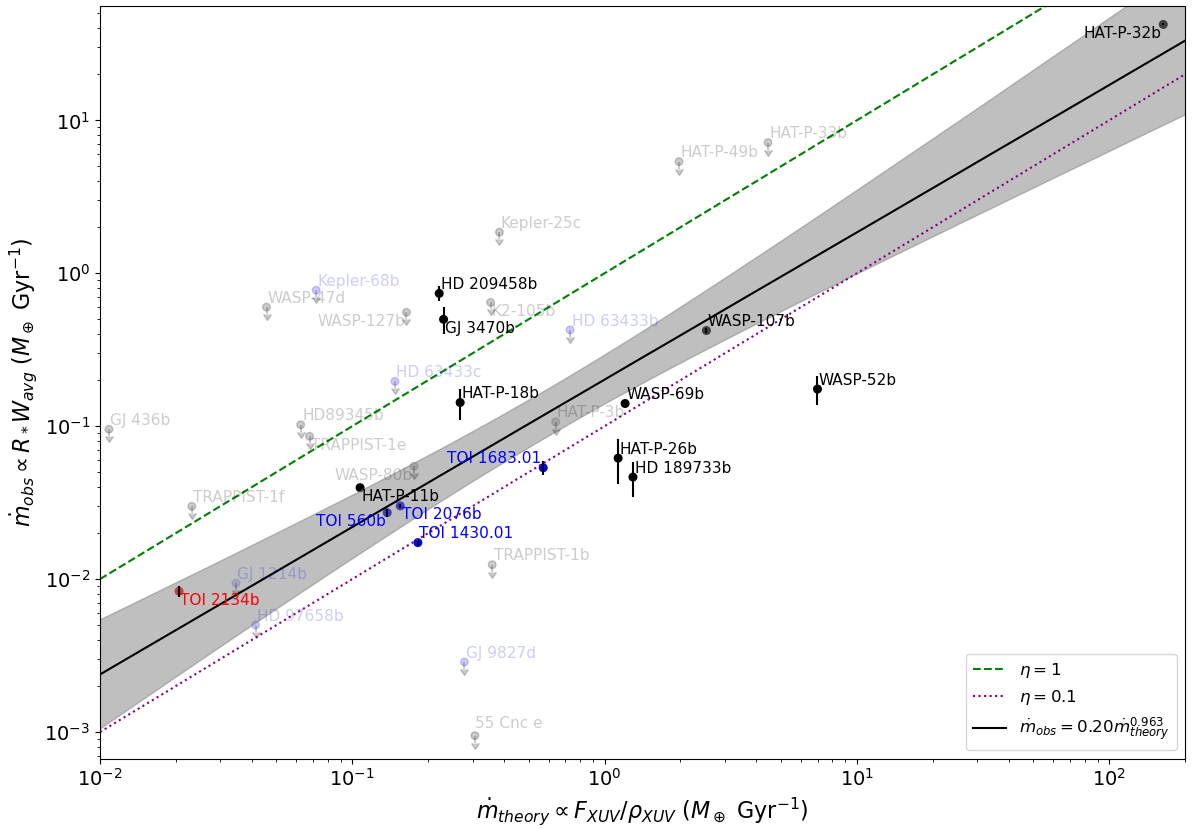}
    \caption{Version of Figure \ref{fig:planets_context} where we attempt to estimate the metastable helium fraction, instead of fixing it at $10^{-6}$.  KELT-9b is far off the top of the chart.}
    \label{fig:planets_context_corr}
\end{figure*}

To test the robustness of the correlation in Figure \ref{fig:planets_context}, we have experimented with other variants.  For example, as mentioned in the main text, we attempt to relax the assumption that $10^{-6}$ of the helium atoms are in the metastable helium state.  \cite{oklopcic_2019} simulated the triplet helium fraction $f$ for a HAT-P-11b-like planet orbiting host stars of 6 different spectral types (their Figure 2), as well as at 6 different semimajor axes from a K1 dwarf (their Figure 5).  For each planet, we first linearly interpolate in $T_{\mathrm{eff}}-\log(f)$ space to obtain an estimate of $f$, and then multiply the estimate by a correction factor obtained from interpolating in $a-\log(f)$ space and dividing by the $f$ at HAT-P-11b's semimajor axis.  For both corrections, we somewhat arbitrarily use the $f$ that \cite{oklopcic_2019} calculates at 3 planetary radii.  Except at a=0.01 AU, $f$ changes very little from 2.5 to 5 $R_p$, and our conclusions do not change when we tried 4 $R_p$ as the reference distance.

Figure \ref{fig:planets_context_corr} shows the relation between $\log(\dot{m}_{\rm theory})$ and $\log(\dot{m}_{\rm obs})$ after these corrections to $f$.  Both \planet and HAT-P-32b now have similar mass loss efficiencies.  The power $p=0.96 \pm 0.18$ is consistent with 1, and the implied typical mass loss efficiency $\eta = 0.20_{-0.06}^{+0.09}$.  On the other hand, the correlation as a whole becomes substantially weaker (ODR p=1.5\e{-4}, Spearman p=0.0054).  Eliminating \planet and HAT-P-32b makes the correlation barely statistically insignificant (ODR p=0.043, Spearman p=0.078).  We conclude that with the possible exception of the two extremes, our crude corrections likely do more harm than good, and that the triplet fraction is unlikely to be a separable function of $T_{\mathrm{eff}}$ and $a$.  Either planet-specific simulations or grid simulations that map out the multidimensional parameter space are likely necessary to obtain accurate estimates of $f$.

Aside from attempting to estimate the metastable fraction, we tried other variants of the correlation.  For example, we tried using the white light radius instead of the estimated XUV radius to estimate the planet density, finding that it weakens the correlation only slightly (ODR p=3.5\e{-5}, Spearman p=2.1\e{-3}).  Dividing the energy-limited mass loss formula by the K-factor, which accounts for the gravitational potential of the star \citep{erkaev_2007}, has a negligible effect on the strength of the correlation (ODR p=2.7\e{-5}, Spearman p=3.8\e{-4}).  Next, we multiplied the energy-limited mass loss rate by an estimate of the efficiency in addition to dividing it by K.  \cite{caldiroli_2022} ran a suite of 1D hydrodynamic simulations using their ATES code and derived an analytic approximation to the mass loss efficiency as a function of $F_{\rm XUV}/\rho$ and the modified gravitational potential $KGM/R$.  After taking this efficiency into account, the correlation slightly strengthens according to the linear regression test (p=9\e{-6}), but slightly weakens according to Spearman's test (p=5.2\e{-4}).

\bibliographystyle{aasjournal} \bibliography{main}
\end{document}